\documentclass[a4paper,12pt,fleqn]{article}%
\usepackage{amsmath, amssymb}
\usepackage{type1cm}
\usepackage{color}
\usepackage{ulem}
\usepackage{amsmath}
\usepackage{amsfonts}
\usepackage{amssymb}
\usepackage{graphicx}%
\providecommand{\U}[1]{\protect\rule{.1in}{.1in}}
\numberwithin{equation}{section}

\begin{document}

\title{\textsf{Runge-Lenz Vector as a 3d Projection of}\\\textsf{ SO(4) Moment Map in} $\mathbb{R}^{4}\times\mathbb{R}^{4}$
\textsf{Phase Space}}
\author{\textsf{Hitoshi Ikemori } \thanks{ikemori@vega.aichi-u.ac.jp}\\Faculty of Economics, Aichi University, \\Nagoya, Aichi 453-8777, Japan
\and \textsf{Shinsaku Kitakado }\thanks{kitakado@ccmfs.meijo-u.ac.jp}\\Professor Emeritus of Nagoya University
\and \textsf{Yoshimitsu Matsui }\thanks{matsui@vega.aichi-u.ac.jp}\\Faculty of Law, Aichi University,\\Nagoya, Aichi 453-8777, Japan
\and \textsf{Toshiro Sato} \thanks{sttshr.aooio@gmail.com}\\School of Engineering, Chukyo University, \\Nagoya, Aichi 466-8666, Japan }
\date{}
\maketitle

\abstract{
We show, using the methods of geometric algebra, that Runge-Lenz vector in the Kepler problem is a 3-dimensional projection of
SO(4) moment map that acts on the phase space of 4-dimensional particle motion.
Thus, RL vector is a consequence of geometric symmetry of $\mathbb{R}^4\times \mathbb{R}^4$ phase
space.
} \noindent

\section{Introduction}

Energy, momentum and angular momentum are the conserved quantities in the
classical mechanics. They take the constant values when considering the motion
of the dynamical system under the influence of certain force. These
quantities, besides being conserved, have another property, i.e. they are the
generators of the group. Momenta, for example, are the generators of the
translation and angular momenta are the generators that induce the rotation
around certain axis. In the framework of symplectic manifolds, this can be
described by the moment map~\cite{Woit}. This map is a generator of a group
$G$ that acts on the phase space and leaves the manifold invariant, in other
words, it is a generator for the infinitesimal canonical transformation.

The definition of the moment map does not imply it to correspond to some
physical quantity, but most of the moment maps represent the conserved
quantities of the dynamical system. Momentum and angular momentum are the
conserved quantities and also the moment maps of the group connected with the
symmetries of the space.

Not all conserved quantities, however, satisfy the conditions for the moment
map. For example Runge-Lenz (RL) vector in the Kepler problem is a typical
case~\cite{RL}. RL vector is known to be conserved in the system, where the
central force is proportional to $1/r^{2}$ .

Let us note here that the phase space for a particle moving in the 2
dimensional Euclidian space can be transformed to any other $2\times2$
dimensional phase space by the so called Levy-Civita mapping~\cite{LC} which
is also a canonical transformation. In particular the 2 dimensional harmonic
oscillator can be transformed to 2 dimensional Kepler system~\cite{Yamamoto}.
Moreover, conserved quantities in the harmonic oscillator are transformed to
part of the RL vector.

Similarly, for 4 dimensional Euclidean space, the phase space is transformed
to the phase space of 3 dimensional particle motion by the
Kustaanheimo-Stiefel (KS) transformation~\cite{KS}. This is the
Marsden-Weinstein theorem~\cite{MW}, which is dimensional reduction of the
phase space by fixing a moment map and keeping the symplectic 2-form. As can
be inferred, 4 dimensional harmonic oscillator can be transformed to the 3
dimensional system of the Kepler problem~\cite{Bartsch,Saha,Iwai}. The
conserved quantities in the 4 dimensional harmonic oscillator system which
generate the SO(4) rotation, satisfy the moment map condition. By the KS
transformation, the moment map is transformed to the RL vector in the 3
dimensional system. Unlike other conserved quantities that are consequences of
the geometric symmetries of the system, conservation of RL vector has been
considered to have its origin in the dynamical properties of the Kepler
problem~\cite{Goldstein}. In this article, we show that RL vector is also a
consequence of geometric symmetry of $\mathbb{R}^{4}\times\mathbb{R}^{4}$
phase space, which manifests itself only for the closed orbits of the
corresponding Kepler system.

In this paper, after giving the definition of the moment map, we show as a
simple example, that Hamiltonian of the harmonic oscillator and angular
momentum are actually the moment maps. We show also that in the 3 dimensional
Kepler system there exists RL vector, which is conserved but apparently is not
a moment map. Next we introduce KS transformation and explain the relation of
4 dimensional harmonic oscillator and 3 dimensional Kepler problem. Finally,
we show that the 4 dimensional momen map can be transformed to 3 dimensional
RL vector of the Kepler system.

\section{Moment Map}

Moment map is defined as follows.

We assume that there is a symplectic manifold $M$, which has a closed and
nondegenerate symplectic 2-form $\omega=dq_{i}\wedge dp_{i}$. Let Lie group
$G$ act on $M$, the map $\mu$,
\begin{align*}
\mu\ :\ M\rightarrow\mathfrak{g}^{*},
\end{align*}
which satisfies the conditions:

\begin{enumerate}
\item For any $x\in M$, $\mu(g\cdot x)=\mathrm{Ad}_{g^{-1}}^{*}\mu(x) \ (g\in
G)$,

\item For a tangent vector field $X$ generated by the action of $G$,
$i_{X}\omega=d\mu$,
\end{enumerate}

is called ``moment map''.

Because the Lie derivative for $\omega$ satisfies
\begin{align*}
\mathcal{L}_{X}\omega=i_{X}d\omega+d(i_{X}\omega),
\end{align*}
the second condition implies that it is a necessary and sufficient condition
for symplectic 2-form $\omega$ to satisfy $\mathcal{L}_{X}\omega=0$, that is,
to guarantee the invariance of $\omega$ in the direction of $X$. Thus, the
moment map $\mu$ is a generator of the canonical transformation.

Suppose $x_{\mu}$ is a coordinate on a certain manifold $M$, then the tangent
vector field $X$ for $x_{\mu}$ is expressed as
\begin{equation}
X={\frac{d}{dt}}=\left.  {\frac{dx_{\mu}}{dt}}\right\vert _{t=0}%
{\frac{\partial}{\partial x_{\mu}}}, \label{tangent vector field}%
\end{equation}
with a parameter $t$. Especially, if the manifold is a phase space
$\mathbb{R}^{n}\times\mathbb{R}^{n}$ consisting of generalized coordinates and
generalized momenta $(q_{i},p_{i})$, then the tangent vector field $X$ is
written as
\[
X=\left.  {\frac{dq_{i}}{dt}}\right\vert _{t=0}{\frac{\partial}{\partial
q_{i}}}+\left.  {\frac{dp_{i}}{dt}}\right\vert _{t=0}{\frac{\partial}{\partial
p_{i}}}.
\]
If there exists a function $h$, satisfying
\[
{\frac{dq_{i}}{dt}}={\frac{\partial h}{\partial p_{i}}},\ {\frac{dp_{i}}{dt}%
}=-{\frac{\partial h}{\partial q_{i}}},
\]
the tangent vector field is called the Hamiltonian vector field $X_{h}$. In
this case, the tangent vector field (\ref{tangent vector field}) becomes
\[
{\frac{d}{dt}}=X_{h}={\frac{\partial h}{\partial p_{i}}}{\frac{\partial
}{\partial q_{i}}}-{\frac{\partial h}{\partial q_{i}}}{\frac{\partial
}{\partial p_{i}}}=-\{h,\cdot\}
\]
It is clear that $h$ is a generator of transformation which acts on the phase
space. If the action is due to the group $G$, $h$ is a moment map defined
above. Therefore, the interior product of $\omega$ and $X_{h}$ becomes
\begin{align*}
i_{X}\omega\equiv\omega(X,\cdot)  &  =dq_{i}\wedge dp_{i}\left(  \left.
{\frac{dq_{i}}{dt}}\right\vert _{t=0}{\frac{\partial}{\partial q_{i}}}+\left.
{\frac{dp_{i}}{dt}}\right\vert _{t=0}{\frac{\partial}{\partial p_{i}}}\right)
\\
&  =dq_{i}\wedge dp_{i}\left(  {\frac{\partial h}{\partial p_{i}}}%
{\frac{\partial}{\partial q_{i}}}-{\frac{\partial h}{\partial q_{i}}}%
{\frac{\partial}{\partial p_{i}}}\right) \\
&  ={\frac{\partial h}{\partial p_{i}}}dp_{i}+{\frac{\partial h}{\partial
q_{i}}}dq_{i}\\
&  =dh.
\end{align*}
This implies that $h$ is a moment map $\mu$ which satisfies the condition (b)
of definition for the moment map. Namely, the action of the group $G$ on the
phase space is symplectomorphism, or canonical transformation.

Now, let us consider a case of U(1) group that acts on $x_{\mu}\equiv
(q_{i},p_{i})\ (i=1,\cdots,n)$ as
\[
x_{\mu}\rightarrow x_{\mu}(t)=(q_{i},p_{i})e^{i\alpha\sigma_{2}t}%
\]
Then, because
\[
\left.  {\frac{dx_{\mu}}{dt}}\right\vert _{t=0}=\alpha(q_{i},p_{i})\left(
\begin{array}
[c]{rr}%
0 & 1\\
-1 & 0
\end{array}
\right)  =\alpha(-p_{i},q_{i}),
\]
tangent vector field is expressed as
\[
X=\left.  {\frac{dx_{\mu}}{dt}}\right\vert _{t=0}{\frac{\partial}{\partial
x_{\mu}}}=\alpha(-p_{i},q_{i})\left(
\begin{array}
[c]{c}%
{{\dfrac{\partial}{\partial q_{i}}}}\\
\\
{{\dfrac{\partial}{\partial p_{i}}}}%
\end{array}
\right)  =\alpha\left(  -{p_{i}}{\frac{\partial}{\partial q_{i}}}+q_{i}%
{\frac{\partial}{\partial p_{i}}}\right)  .
\]
Substituting $X$ into the relation $\omega(X,\cdot)=dh$, we obtain
\[
p_{i}dp_{i}+q_{i}dq_{i}=dh.
\]
Therefore,
\[
h={\frac{1}{2}}(p_{i}^{2}+q_{i}^{2}).
\]
Thus, $h$ is a moment map. Therefore, the Hamiltonian of harmonic oscillator
is moment map by the action of U(1) group on phase space.

Next, we show that angular momentum is another typical example of the moment
maps. Let us consider a SO(3) group acting on the coordinate of the manifold
$\mathbb{R}^{3}\times\mathbb{R}^{3}$. This group action is expressed by the
Pauli matrices as
\[
(q_{i},p_{i})\sigma_{i}=(q_{i}\sigma_{i},\ p_{i}\sigma_{i})\rightarrow
e^{i\alpha_{j}\sigma_{j}t}(q_{i}\sigma_{i},\ p_{i}\sigma_{i})e^{-i\alpha
_{j}\sigma_{j}t}.
\]
Then,
\begin{align*}
&  {}\left.  {\frac{d}{dt}}\left(  e^{i\alpha_{j}\sigma_{j}t}q_{i}\sigma
_{i}e^{-i\alpha_{j}\sigma_{j}t}\right)  \right\vert _{t=0}=i\alpha_{i}%
q_{j}[\sigma_{i},\sigma_{j}]=-2\epsilon_{ijk}\alpha_{i}q_{j}\sigma_{k}\\
&  {}\left.  {\frac{d}{dt}}\left(  e^{i\alpha_{j}\sigma_{j}t}p_{i}\sigma
_{i}e^{-i\alpha_{j}\sigma_{j}t}\right)  \right\vert _{t=0}=i\alpha_{i}%
p_{j}[\sigma_{i},\sigma_{j}]=-2\epsilon_{ijk}\alpha_{i}p_{j}\sigma_{k}.
\end{align*}
From these equations, the tangent vector field is obtained as
\[
X=\left.  {\frac{dx_{\mu}}{dt}}\right\vert _{t=0}{\frac{\partial}{\partial
x_{\mu}}}=-2\alpha_{i}(\epsilon_{ijk}q_{j},\epsilon_{ijk}p_{j})\left(
\begin{array}
[c]{c}%
{{\dfrac{\partial}{\partial q_{k}}}}\\
\\
{{\dfrac{\partial}{\partial p_{k}}}}%
\end{array}
\right)  =-2\alpha_{i}\left(  \epsilon_{ijk}q_{j}{\frac{\partial}{\partial
q_{k}}}+\epsilon_{ijk}p_{j}{\frac{\partial}{\partial p_{k}}}\right)  .
\]
Interior product of $\omega$ and $X$ gives
\[
\omega(X,\cdot)=(\epsilon_{ijk}p_{j}dq_{k}-\epsilon_{ijk}q_{j}dp_{k}%
)\alpha_{i}=d(\epsilon_{ijk}q_{j}p_{k})\alpha_{i},
\]
and the moment map is expressed as
\begin{equation}
\mu_{i}=\epsilon_{ijk}q_{j}p_{k}. \label{mui}%
\end{equation}
Since the right hand side of this expression is the angular momentum, this
means that the angular momentum is the momemt map of SO(3) group acting on the
phase space $\mathbb{R}^{3}\times\mathbb{R}^{3}$.

\section{Kepler Problem and Runge-Lenz Vector}

It is well known that there exists a conserved quantity called Runge-Lenz (RL)
vector~\cite{RL} in the dynamical system of the Kepler problem, apart from
Hamiltonian and angular momentum.

In the 3 dimensional dynamical system of the Kepler problem, equation of
motion is
\begin{equation}
\boldsymbol{\ddot{r}}=-{\frac{k}{r^{3}}}\boldsymbol{r},
\end{equation}
where $\boldsymbol{r}=(x_{1},x_{2},x_{3})$ is the coordinate vector of a
particle.
\[
\boldsymbol{\dot{L}}={\frac{d}{dt}}(\boldsymbol{r}\times\boldsymbol{\dot{r}%
})=0,
\]
since the force of the system is central.
\[
\boldsymbol{\ddot{r}}\times\boldsymbol{L}=\boldsymbol{\ddot{r}}\times
(\boldsymbol{r}\times\boldsymbol{\dot{r}})=-{\frac{k}{r^{3}}}\boldsymbol{r}%
\times(\boldsymbol{r}\times\boldsymbol{\dot{r}}),
\]
and using the relation
\[
{\frac{d}{dt}}\{\boldsymbol{\dot{r}}\times(\boldsymbol{r}\times
\boldsymbol{\dot{r}})\}=\boldsymbol{\ddot{r}}\times(\boldsymbol{r}%
\times\boldsymbol{\dot{r}})+\boldsymbol{\dot{r}}\times(\boldsymbol{\dot{r}%
}\times\boldsymbol{\dot{r}})+\boldsymbol{\dot{r}}\times(\boldsymbol{r}%
\times\boldsymbol{\ddot{r}})=\boldsymbol{\ddot{r}}\times(\boldsymbol{r}%
\times\boldsymbol{\dot{r}}),
\]
the equation of motion becomes
\[
{\frac{d}{dt}}\{\boldsymbol{\dot{r}}\times(\boldsymbol{r}\times
\boldsymbol{\dot{r}})\}=-{\frac{k}{r^{3}}}\boldsymbol{r}\times(\boldsymbol{r}%
\times\boldsymbol{\dot{r}}).
\]
Moreover, since the right hand side of this equation can be rewritten as
\[
{\frac{\boldsymbol{r}\times(\boldsymbol{r}\times\boldsymbol{\dot{r}})}{r^{3}}%
}={\frac{(\boldsymbol{r}\cdot\boldsymbol{\dot{r}})\boldsymbol{r}%
-r^{2}\boldsymbol{\dot{r}}}{r^{3}}}=-{\frac{d}{dt}}{\frac{\boldsymbol{r}}{r}%
},
\]
then the equation of motion becomes
\[
{\frac{d}{dt}}\{\boldsymbol{\dot{r}}\times(\boldsymbol{r}\times
\boldsymbol{\dot{r}})\}=k{\frac{d}{dt}}{\frac{\boldsymbol{r}}{r}}.
\]
Thus,
\[
{\frac{d}{dt}}\left\{  \boldsymbol{\dot{r}}\times(\boldsymbol{r}%
\times\boldsymbol{\dot{r}})-k{\frac{\boldsymbol{r}}{r}}\right\}  =0.
\]
This implies that the content in the parentheses is constant of motion. This
quantity is called the RL vector $\boldsymbol{A}$. In this discussion, the
potential term is necessary to derive the RL vector. Therefore, RL vector is a
conserved quantity which depends on the dynamical system of the Kepler
problem. Thus, it seems that RL vector has nothing to do with moment map which
depends on geometry of the phase space. However, if motion of the particle
draws an elliptic orbit, i.e., the value of dynamical energy $E$ is negative,
then together with the moment map $\mu_{i}$, (\ref{mui}), the normalized RL
vector $\tilde{A}_{i}=A_{i}/\sqrt{-2E}$ satisfies the SO(4) algebra,
\begin{align}
&  {}\{\mu_{i},\mu_{j}\}=\epsilon_{ijk}\mu_{k},\label{AA}\\
&  {}\{\mu_{i},\tilde{A}_{j}\}=\epsilon_{ijk}\tilde{A}_{k},\\
&  {}\{\tilde{A}_{i},\tilde{A}_{j}\}=\epsilon_{ijk}\mu_{k}.
\end{align}
This suggests that the RL vector is related to the moment map.

\section{Conserved Quantity in 4-Dimensional Harmonic Oscillator and SO(4)
Moment Map}

It is necessary to find the SO(4) moment map of the $\mathbb{R}^{4}%
\times\mathbb{R}^{4}$ phase space in order to see the relation between the RL
vector of the Kepler problem in the three dimensional space and the moment map
of the $\mathbb{R}^{4}\times\mathbb{R}^{4}$ phase space.

In the previous discussion, we found out that the moment map in the phase
space is related to the conserved quantity in the harmonic oscillator system.
Therefore, it is expected that appropriate combinations of conserved
quantities in the four dimensional harmonic oscillator constitute the
generators of the SO(4) group, and become the SO(4) moment maps. In the four
dimensional harmonic oscillator, the Hamiltonian
\begin{equation}
H=\sum_{\alpha=0}^{3}\left(  {\frac{1}{2}}p_{\alpha}^{2}+{\frac{1}{2}%
}q_{\alpha}^{2}\right)  ,
\end{equation}
and the angular momenta
\begin{equation}
L_{i}^{L(R)}=q_{i}p_{0}-q_{0}p_{i}\pm\epsilon_{ijk}q_{j}p_{k}\quad(i=1,2,3),
\label{angulthegeneratorsar momentum}%
\end{equation}
are the typical conserved quantities in the system. However, there are the
other conserved quantities, such as~\cite{Katayama}
\begin{equation}
J_{\alpha\beta}=p_{\alpha}p_{\beta}+q_{\alpha}q_{\beta}.\quad(\alpha
,\beta=0,1,2,3)
\end{equation}
Using these quantities, let us consider the following combinations,
\begin{align}
K_{1}  &  ={\frac{1}{2}}(J_{13}-J_{02}),\label{K_1}\\
K_{2}  &  ={\frac{1}{2}}(J_{01}+J_{23}),\label{K_2}\\
K_{3}  &  ={\frac{1}{4}}(J_{00}+J_{33}-J_{11}-J_{22}). \label{K_3}%
\end{align}
Moreover, it is easy to find that they satisfy the SO(4) algebra, together
with a part of angular momentum $L_{i}^{L}$,
\begin{align}
&  \{L_{i}^{L},L_{j}^{L}\}=\epsilon_{ijk}L_{k}^{L}\label{KK}\\
&  \{L_{i}^{L},K_{j}\}=\epsilon_{ijk}K_{k}\\
&  \{K_{i},K_{j}\}=\epsilon_{ijk}L_{k}^{L}%
\end{align}
As shown in the appendix, $K_{i}$'s and $L_{i}^{L}$'s generate rotations in
the phase space.

These generators are represented by the matrices $\Sigma_{i},\ \Lambda_{i}$.
For example, the matrix, which corresponds to the infinitesimal rotation
generated by $K_{1}$, is
\begin{align*}
\Sigma_{1}={\frac{1}{2}}\left(
\begin{array}
[c]{rrrr|rrrr}%
0 & 0 & 0 & 0 & 0 & 0 & 1 & 0\\
0 & 0 & 0 & 0 & 0 & 0 & 0 & -1\\
0 & 0 & 0 & 0 & 1 & 0 & 0 & 0\\
0 & 0 & 0 & 0 & 0 & -1 & 0 & 0\\\hline
0 & 0 & -1 & 0 & 0 & 0 & 0 & 0\\
0 & 0 & 0 & 1 & 0 & 0 & 0 & 0\\
-1 & 0 & 0 & 0 & 0 & 0 & 0 & 0\\
0 & 1 & 0 & 0 & 0 & 0 & 0 & 0
\end{array}
\right)  .
\end{align*}
Indeed, acting this matrix on the phase space, we obtain
\begin{align*}
\Sigma_{1}\left(
\begin{array}
[c]{c}%
Q\\
P
\end{array}
\right)  ={\frac{1}{2}}\left(
\begin{array}
[c]{rrrr|rrrr}%
0 & 0 & 0 & 0 & 0 & 0 & 1 & 0\\
0 & 0 & 0 & 0 & 0 & 0 & 0 & -1\\
0 & 0 & 0 & 0 & 1 & 0 & 0 & 0\\
0 & 0 & 0 & 0 & 0 & -1 & 0 & 0\\\hline
0 & 0 & -1 & 0 & 0 & 0 & 0 & 0\\
0 & 0 & 0 & 1 & 0 & 0 & 0 & 0\\
-1 & 0 & 0 & 0 & 0 & 0 & 0 & 0\\
0 & 1 & 0 & 0 & 0 & 0 & 0 & 0
\end{array}
\right)  \left(
\begin{array}
[c]{c}%
q_{0}\\
q_{1}\\
q_{2}\\
q_{3}\\
p_{0}\\
p_{1}\\
p_{2}\\
p_{3}%
\end{array}
\right)  ={\frac{1}{2}} \left(
\begin{array}
[c]{r}%
p_{2}\\
-p_{3}\\
p_{0}\\
-p_{1}\\
-q_{2}\\
q_{3}\\
-q_{0}\\
q_{1}%
\end{array}
\right)  .
\end{align*}
We find that this matrix generates the infinitesimal rotation which is caused
by $K_{1}$. (About the explicit form of the other matrices, see Appendix A)
The matrices $\Sigma_{i},\Lambda_{i}$ also satisfy the SO(4) algebra,
\begin{align*}
&  \{i\Lambda_{i},i\Lambda_{j}\}=\epsilon_{ijk}i\Lambda_{k}\\
&  \{i\Lambda_{i},i\Sigma_{j}\}=\epsilon_{ijk}i\Sigma_{k}\\
&  \{i\Sigma_{i},i\Sigma_{j}\}=\epsilon_{ijk}i\Lambda.
\end{align*}

Introducing the matrices $\Omega^{L}\ $and $\Omega^{R}$ which are defined as
$\Omega_{i}^{L}={\dfrac{\Lambda_{i}-\Sigma_{i}}{2}}$ and $\ \Omega_{i}%
^{R}={\dfrac{\Lambda_{i}+\Sigma_{i}}{2}}$, these matrices satisfy
SO(3)$_{\mathrm{L}}$ and SO(3)$_{\mathrm{R}}$ algebras, independently.

Using these matrices, we can construct tangent vector field connected with
$\Omega_{i}^{L(R)}$. For example, the tangent vector field for infinitesimal
rotation generated by $\Omega_{i}^{L}$ can be obtained. We define $A$ as
\begin{equation}
A=e^{i\Omega_{a}^{L}\alpha_{a}t}\left(
\begin{array}
[c]{c}%
Q\\
P
\end{array}
\right)  \simeq(1+i\Omega_{a}^{L}\alpha_{a}t)\left(
\begin{array}
[c]{c}%
Q\\
P
\end{array}
\right)  , \label{AL}%
\end{equation}
where $(Q,P)^{T}$ is the coordinates of phase space, defined as
\begin{align*}
Q  &  =\left(
\begin{array}
[c]{c}%
Q_{1}\\
Q_{2}%
\end{array}
\right)  ,\ P=\left(
\begin{array}
[c]{c}%
P_{1}\\
P_{2}%
\end{array}
\right)  ,\\
\ Q_{1}  &  =\left(
\begin{array}
[c]{c}%
q_{0}\\
q_{1}%
\end{array}
\right)  ,\ Q_{2}=\left(
\begin{array}
[c]{c}%
q_{2}\\
q_{3}%
\end{array}
\right)  ,\\
\ P_{1}  &  =\left(
\begin{array}
[c]{c}%
p_{0}\\
p_{1}%
\end{array}
\right)  ,\ P_{2}=\left(
\begin{array}
[c]{c}%
p_{2}\\
p_{3}%
\end{array}
\right)  .
\end{align*}
From (\ref{AL}), the tangent vector field becomes
\begin{equation}
X={\frac{d}{dt}}=\left.  {\frac{dA^{T}}{dt}}\right\vert _{t=0}{\frac{\partial
}{\partial A}}=i(Q^{T},P^{T})(\Omega_{a}^{L})^{T}\alpha_{a}\left(
\begin{array}
[c]{c}%
\partial_{Q}\\
\partial_{P}%
\end{array}
\right)  , \label{X}%
\end{equation}
where
\begin{align*}
{\frac{\partial}{\partial A}}  &  =\left(
\begin{array}
[c]{c}%
\partial_{Q}\\
\partial_{P}%
\end{array}
\right)  ,\\
\partial_{Q}  &  =\left(
\begin{array}
[c]{c}%
\partial_{Q_{1}}\\
\partial_{Q_{2}}%
\end{array}
\right)  ,\ \partial_{Q_{1}}=\left(
\begin{array}
[c]{c}%
\partial_{q_{0}}\\
\partial_{q_{1}}%
\end{array}
\right)  ,\ \partial_{Q_{2}}=\left(
\begin{array}
[c]{c}%
\partial_{q_{2}}\\
\partial_{q_{3}}%
\end{array}
\right)  ,\\
\partial_{P}  &  =\left(
\begin{array}
[c]{c}%
\partial_{P_{1}}\\
\partial_{P_{2}}%
\end{array}
\right)  ,\ \partial_{P_{1}}=\left(
\begin{array}
[c]{c}%
\partial_{p_{0}}\\
\partial_{p_{1}}%
\end{array}
\right)  ,\ \partial_{P_{2}}=\left(
\begin{array}
[c]{c}%
\partial_{p_{2}}\\
\partial_{p_{3}}%
\end{array}
\right)
\end{align*}
Interior product of the tangent vector field (\ref{X}) and the symplectic
2-form $\omega=dq_{\alpha}\wedge dp_{\alpha}$ becomes
\begin{equation}
\omega(X,\cdot)=d(L_{a}^{L}-K_{a})\alpha_{a}. \label{omega}%
\end{equation}
This implies that the $L_{a}^{L}-K_{a}$ is the SO(3)$_{\mathrm{L}}$ moment
map. In the same manner, using $\Omega_{i}^{R}$ in substitution for
$\Omega_{i}^{L}$, it can be shown that $L_{a}^{L}+K_{a}$ is the
SO(3)$_{\mathrm{R}}$ moment map. Therefore, the $L_{a}^{L}\pm K_{a}$ are the
SO(4) moment maps in the $\mathbb{R}^{4}\times\mathbb{R}^{4}$ phase space.

\section{Kustaanheimo-Stiefel Transformation}

In order to investigate the relation between the RL vector and the SO(4)
moment map in 4 dimensional system, we have to use the symplectic reduction
and Marsden \& Weinstein (MW) theorem~\cite{MW}.

MW theorem and symplectic quotient are defined as follows:

\begin{itemize}
\item Orbital space $M_{\mathrm{red}}=\mu^{-1}(0)/G$ is manifold.

\item Map $\pi:\mu^{-1}(0)\rightarrow M_{\mathrm{red}}$ is principal $G$-bundle.

\item There exists symplectic 2-form $\omega_{\mathrm{red}}$, satisfying
$i^{*}\omega=\pi^{*}\omega_{\mathrm{red}}$ in the $M_{\mathrm{red}}$.
\end{itemize}

$(M_{\mathrm{red}},\omega_{\mathrm{red}})$ is called the symplectic quotient
for $(M,\omega)$ about $G$. In other words, the MW theorem is dimensional
reduction of the phase space by fixing a moment map, keeping symplectic 2-form.

Especially, fixing a U(1) moment map, the symplectic reduction from the
4-dimensional space to 3-dimensional space is called the Kustaanheimo-Stiefel
(KS) transformation~\cite{KS}.

We use the geometric algebra to represent the KS transformation, in the
following discussion. The geometric algebra is a method to express a system
using the bases $\boldsymbol{\sigma}_{i}$, which satisfy the relations
\begin{equation}
\boldsymbol{\sigma}_{i}\boldsymbol{\sigma}_{j}+\boldsymbol{\sigma}%
_{j}\boldsymbol{\sigma}_{i}=2\delta_{ij}.\label{GA}%
\end{equation}
Then, the generalized coordinates and momenta in 4-dimensional space are
expressed as
\begin{align*}
Q &  =q_{0}+q_{1}\boldsymbol{\sigma}_{2}\boldsymbol{\sigma}_{3}+q_{2}%
\boldsymbol{\sigma}_{3}\boldsymbol{\sigma}_{1}+q_{3}\boldsymbol{\sigma}%
_{1}\boldsymbol{\sigma}_{2},\\
P &  =p_{0}-p_{1}\boldsymbol{\sigma}_{2}\boldsymbol{\sigma}_{3}-p_{2}%
\boldsymbol{\sigma}_{3}\boldsymbol{\sigma}_{1}-p_{3}\boldsymbol{\sigma}%
_{1}\boldsymbol{\sigma}_{2}.
\end{align*}
From these definition, the generalized coordinate $Q$ satisfies the relation
\begin{equation}
QQ^{\dagger}=Q^{\dagger}Q=2(q_{0}^{2}+q_{1}^{2}+q_{2}^{2}+q_{3}^{2})\equiv2r.
\end{equation}
The KS transformation in terms of the geometric algebra is expressed as
\begin{equation}
\boldsymbol{x}={\frac{1}{2}}Q\boldsymbol{\sigma}_{3}Q^{\dagger},\label{x}%
\end{equation}
where $\boldsymbol{x}$ is the space coordinate in 3-dimensional space, and is
expressed as
\begin{align}
\boldsymbol{x} &  \equiv x_{1}\boldsymbol{\sigma}_{1}+x_{2}\boldsymbol{\sigma
}_{2}+x_{3}\boldsymbol{\sigma}_{3}\nonumber\\
&  =(q_{1}q_{3}-q_{0}q_{2})\boldsymbol{\sigma}_{1}+(q_{1}q_{0}+q_{2}%
q_{3})\boldsymbol{\sigma}_{2}+{\frac{1}{2}}(q_{0}^{2}+q_{3}^{2}-q_{1}%
^{2}-q_{2}^{2})\boldsymbol{\sigma}_{3},\nonumber\\
\boldsymbol{x}^{2} &  =x_{1}^{2}+x_{2}^{2}+x_{3}^{2}=r^{2},\label{x^2}%
\end{align}
so that time derivatives of $x_{i}$'s become
\begin{align*}
\dot{x}_{1} &  =\dot{q}_{1}q_{3}+q_{1}\dot{q}_{3}-\dot{q}_{0}q_{2}-q_{0}%
\dot{q}_{2},\\
\dot{x}_{2} &  =\dot{q}_{0}q_{1}+q_{0}\dot{q}_{1}+\dot{q}_{2}q_{3}+q_{2}%
\dot{q}_{3},\\
\dot{x}_{3} &  =\dot{q}_{0}q_{0}+\dot{q}_{3}q_{3}-\dot{q}_{1}q_{1}-\dot{q}%
_{2}q_{2}.
\end{align*}
Here, we define
\begin{equation}
{\frac{\mu}{r}}\equiv\dot{q}_{0}q_{3}-q_{0}\dot{q}_{3}+\dot{q}_{1}q_{2}%
-q_{1}\dot{q}_{2},\label{y}%
\end{equation}
where $\mu$ corresponds to the righthanded angular momentum $L_{3}^{R}$ in
(\ref{angulthegeneratorsar momentum}), and is the U(1) moment map which is
fixed to reduce a degree of freedom of the phase space. When we assume $\mu
=0$,
\[
\dot{x}_{1}^{2}+\dot{x}_{2}^{2}+\dot{x}_{3}^{2}=r(\dot{q}_{0}^{2}+\dot{q}%
_{1}^{2}+\dot{q}_{2}^{2}+\dot{q}_{3}^{2}).
\]
After all, Lagrangian in the 3-dimensional dynamical system
\[
L={\frac{1}{2}}(\dot{x}_{1}^{2}+\dot{x}_{2}^{2}+\dot{x}_{3}^{2}+\mu
^{2})-V(x_{i})
\]
can be rewritten in 4-dimensional system as
\[
L={\frac{1}{2}}r(\dot{q}_{0}^{2}+\dot{q}_{1}^{2}+\dot{q}_{2}^{2}+\dot{q}%
_{3}^{2})-V(q_{\mu}),
\]
where $V$ is a certain potential term depending on $x_{i}$ or $q_{\mu}$. From
this Lagrangian, we obtain the generalized momenta in 4-dimensional
coordinates, such as
\[
p_{\mu}\equiv{\frac{\partial L}{\partial\dot{q}_{\mu}}}=r\dot{q}_{\mu}.
\]
This implies that, in the geometric algebra, 3-dimensional momentum
$\boldsymbol{\pi}$ can be written as
\begin{equation}
\boldsymbol{\pi}={\frac{1}{r}}\langle Q\boldsymbol{\sigma}_{3}P\rangle_{1},
\end{equation}
where $\langle A\rangle_{i}$ represents the part of grade-$i$ (see Appendix B).

It is known that, using the KS reduction, the dynamical system of the harmonic
oscillator in 4-dimensional space can be regarded as that of the Kepler
problem in 3-dimensional space. This fact is easily shown by use of the
geometric algebra.

In the geometric algebra, the Hamiltonian of the harmonic oscillator is
represented by
\begin{align*}
H={\frac{1}{2}}P^{\dagger}P+{\frac{\lambda^{2}}{2}}QQ^{\dagger}.
\end{align*}
We note that we have introduced the coefficient of the potential term
$\lambda$. The $\lambda$ plays an important role in the subsequent discussion.

Using some properties of $Q,P$, Hamiltonian is rewritten as
\[
H={\frac{1}{2r}}\left(  {\frac{1}{2}}P^{\dagger}{\boldsymbol{\sigma}}%
_{3}Q^{\dagger}Q{\boldsymbol{\sigma}}_{3}P+{\frac{\lambda^{2}}{2}%
}Q{\boldsymbol{\sigma}}_{3}Q^{\dagger}Q{\boldsymbol{\sigma}}_{3}Q^{\dagger
}\right)  .
\]
Moreover, from (\ref{x}),(\ref{x^2}),(\ref{y}) and
\begin{align*}
&  \frac{1}{r}Q{\boldsymbol{\sigma}}_{3}P=\pi_{1}{\boldsymbol{\sigma}}_{1}%
+\pi_{2}{\boldsymbol{\sigma}}_{2}+\pi_{3}{\boldsymbol{\sigma}}_{3}%
+{\mu\boldsymbol{\sigma}}_{1}{\boldsymbol{\sigma}}_{2}{\boldsymbol{\sigma}%
}_{3},\\
&  \frac{1}{r}P^{\dagger}{\boldsymbol{\sigma}}_{3}Q^{\dagger}=\pi
_{1}{\boldsymbol{\sigma}}_{1}+\pi_{2}{\boldsymbol{\sigma}}_{2}+\pi
_{3}{\boldsymbol{\sigma}}_{3}-{\mu\boldsymbol{\sigma}}_{1}{\boldsymbol{\sigma
}}_{2}{\boldsymbol{\sigma}}_{3},
\end{align*}
Hamiltonian becomes
\begin{equation}
H={\frac{r}{4}\boldsymbol{\pi}}^{2}+\lambda^{2}r,\label{H2}%
\end{equation}
when $\mu=0$. If we replace $\lambda^{2}$ with $-{\dfrac{h}{2}}$, and $H$ with
${\dfrac{k}{2}}$, the following expression is obtained except for $r=0$, when
the both sides of (\ref{H2}) are divided by $r$.
\begin{equation}
h={\frac{1}{2}\boldsymbol{\pi}}^{2}-{\frac{k}{r}}.\label{Kepler}%
\end{equation}
This is interpreted as the Hamiltonian of the Kepler problem in 3 dimensional space.

Of course, $h$ is not the Hamiltonian of the original harmonic oscillator
system in 4 dimensional space, but is the coefficient of the potential term.
Since a role of the Hamiltonian is replaced with the coefficient of the
potential term, these dynamical systems are not equivalent to each other.
Furthermore, (\ref{Kepler}) can be interpreted as the Hamiltonian of the
Kepler problem, only if $h=-2\lambda^{2}<0$. The KS transformation is one of
the canonical transformation from $\mathbb{R}^{4}\times\mathbb{R}^{4}$ to
$\mathbb{R}^{3}\times\mathbb{R}^{3}$, since it is the transformation which
keeps the symplectic 2-form invariant. Therefore, as far as we consider the
dynamics of closed orbit with a fixed value of negative energy in the 3
dimensional Kepler problem, we can treat it as the dynamical motion in the 4
dimensional harmonic oscillator, regarding $h$ as the coefficient of the
potential term.

\section{Runge-Lenz Vector as Moment Map}

In the previous section, it was shown that, using the geometric algebra, the
dynamical system of the 4 dimensional harmonic oscillator and that of the 3
dimensional Kepler problem were related to each other through the KS
transformation. Therefore, it can be expected that we are able to show the
relation between the RL vector in 3 dimensional Kepler problem and the moment
map in the phase space of 4 dimensional particle motion, using the geometric algebra.

Let us consider the linear combination of the conserved quantities of the 4
dimensional harmonic oscillator,
\begin{align}
&  P^{\dagger}\boldsymbol{\sigma}_{3}P+\lambda^{2}Q\boldsymbol{\sigma}%
_{3}Q^{\dagger}\nonumber\\
&  =\left\{  (p_{1}p_{3}-p_{0}p_{2})+\lambda^{2}(q_{1}q_{3}-q_{0}%
q_{2})\right\}  \boldsymbol{\sigma}_{1}\nonumber\\
&  +\left\{  (p_{0}p_{1}+p_{2}p_{3})+\lambda^{2}(q_{0}q_{1}+q_{2}%
q_{3})\right\}  \boldsymbol{\sigma}_{2}\nonumber\\
&  +\left\{  {\frac{1}{2}}(p_{0}^{2}+p_{3}^{2}-p_{1}^{2}-p_{2}^{2})+{\frac
{1}{2}}\lambda^{2}(q_{0}^{2}+q_{3}^{2}-q_{1}^{2}-q_{2}^{2})\right\}
\boldsymbol{\sigma}_{3}\nonumber\\
&  \equiv(J_{13}-J_{02})\boldsymbol{\sigma}_{1}+(J_{01}+J_{23}%
)\boldsymbol{\sigma}_{2}+{\frac{1}{2}}(J_{00}+J_{33}-J_{11}-J_{22}%
)\boldsymbol{\sigma}_{3} \label{JJ}%
\end{align}
This is the SO(4) moment map (\ref{K_1}),(\ref{K_2}),(\ref{K_3}) in the phase
space of the 4 dimensional particle motion represented by means of the method
of the geometric algebra.\footnote{In (\ref{JJ}), we assume $\mu=0$ and
$\lambda\neq1$.}. It is obvious that the relation
\begin{equation}
Q\boldsymbol{\sigma}_{3}P=P^{\dagger}\boldsymbol{\sigma}_{3}Q^{\dagger},
\label{QP=PQ}%
\end{equation}
is satisfied, since
\begin{align*}
Q\boldsymbol{\sigma}_{3}P=  &  (q_{0}+q_{1}\boldsymbol{\sigma}_{2}%
\boldsymbol{\sigma}_{3}+q_{2}\boldsymbol{\sigma}_{3}\boldsymbol{\sigma}%
_{1}+q_{3}\boldsymbol{\sigma}_{1}\boldsymbol{\sigma}_{2})\boldsymbol{\sigma
}_{3}(p_{0}-p_{1}\boldsymbol{\sigma}_{2}\boldsymbol{\sigma}_{3}-p_{2}%
\boldsymbol{\sigma}_{3}\boldsymbol{\sigma}_{1}-p_{3}\boldsymbol{\sigma}%
_{1}\boldsymbol{\sigma}_{2})\\
=  &  (q_{0}\boldsymbol{\sigma}_{3}+q_{1}\boldsymbol{\sigma}_{2}%
-q_{2}\boldsymbol{\sigma}_{1}+q_{3}I)\boldsymbol{\sigma}_{3}(p_{0}%
-p_{1}\boldsymbol{\sigma}_{2}\boldsymbol{\sigma}_{3}-p_{2}\boldsymbol{\sigma
}_{3}\boldsymbol{\sigma}_{1}-p_{3}\boldsymbol{\sigma}_{1}\boldsymbol{\sigma
}_{2})\\
=  &  (q_{1}p_{3}+q_{3}p_{1}-q_{0}p_{2}-q_{2}p_{0})\boldsymbol{\sigma}%
_{1}+(q_{0}p_{1}+q_{1}p_{0}+q_{2}p_{3}+q_{3}p_{2})\boldsymbol{\sigma}_{2}\\
&  +(q_{0}p_{0}+q_{3}p_{3}-q_{1}p_{1}-q_{2}p_{2})\boldsymbol{\sigma}%
_{3}+(q_{3}p_{0}-q_{0}p_{3}+q_{2}p_{1}-q_{1}p_{2})\boldsymbol{\sigma}%
_{1}\boldsymbol{\sigma}_{2}\boldsymbol{\sigma}_{3},
\end{align*}
and ${\mu=}\langle Q\boldsymbol{\sigma}_{3}P\rangle_{3}=0$ is assumed.

Based on (\ref{QP=PQ}), we can deform (\ref{JJ}) as
\begin{align}
&  P^{\dagger}\boldsymbol{\sigma}_{3}P+\lambda^{2}Q\boldsymbol{\sigma}%
_{3}Q^{\dagger}\nonumber\\
&  =P^{\dagger}\boldsymbol{\sigma}_{3}P-{\frac{1}{2}}\left(  {\frac{1}%
{2}\boldsymbol{\pi}}^{2}-{\frac{k}{r}}\right)  (Q\boldsymbol{\sigma}%
_{3}Q^{\dagger})\nonumber\\
&  =P^{\dagger}\boldsymbol{\sigma}_{3}P-{\frac{1}{4r^{2}}}P^{\dagger
}\boldsymbol{\sigma}_{3}Q^{\dagger}P^{\dagger}\boldsymbol{\sigma}%
_{3}Q^{\dagger}Q\boldsymbol{\sigma}_{3}Q^{\dagger}+{\frac{k}{2r}%
}Q\boldsymbol{\sigma}_{3}Q^{\dagger}\nonumber\\
&  =P^{\dagger}\boldsymbol{\sigma}_{3}P-{\frac{1}{2r}}P^{\dagger
}\boldsymbol{\sigma}_{3}Q^{\dagger}P^{\dagger}Q^{\dagger}+{k}%
Q\boldsymbol{\sigma}_{3}Q^{-1}, \label{Ps3P+Qs3Q}%
\end{align}
where we use the fact that $\lambda^{2}$ is Hamiltonian of the Kepler problem
in 3-dimsnsional space.

On the other hand, RL vector $A\ (=A_{1}\boldsymbol{\sigma}_{1}+A_{2}%
\boldsymbol{\sigma}_{2}+A_{3}\boldsymbol{\sigma}_{3})$ is written as
\[
A=l\boldsymbol{\dot{x}}-k{\frac{\boldsymbol{x}}{r}},
\]
in the geometric algebra. Since it is known that the 3 dimensional angular
momentum $l$ and $\boldsymbol{\dot{x}}$ are expressed as\footnote{We note that
$l$ is exactly the same as the 4 dimensional angular momentum $L^{L}=\langle
QP\rangle_{2}$}
\[
l=\langle\boldsymbol{x\dot{x}}\rangle_{2}=\langle QP\rangle_{2}={\frac{1}{2}%
}\left(  QP-P^{\dagger}Q^{\dagger}\right)  ,
\]
and
\[
\boldsymbol{\dot{x}}=\dot{Q}\boldsymbol{\sigma}_{3}Q^{\dagger}=2P^{\dagger
}\boldsymbol{\sigma}_{3}Q^{-1},
\]
the RL vector is rewritten as
\begin{align}
A  &  =2lP^{\dagger}\boldsymbol{\sigma}_{3}Q^{-1}-kQ\boldsymbol{\sigma}%
_{3}Q^{-1}\nonumber\\
&  =2\langle QP\rangle_{2}P^{\dagger}\boldsymbol{\sigma}_{3}Q^{-1}%
-kQ\boldsymbol{\sigma}_{3}Q^{-1}\nonumber\\
&  =(QP-P^{\dagger}Q^{\dagger})P^{\dagger}\boldsymbol{\sigma}_{3}%
Q^{-1}-kQ\boldsymbol{\sigma}_{3}Q^{-1}\nonumber\\
&  ={\frac{1}{2r}}QPP^{\dagger}\boldsymbol{\sigma}_{3}Q^{\dagger}-{\frac
{1}{2r}}P^{\dagger}Q^{\dagger}P^{\dagger}\boldsymbol{\sigma}_{3}Q^{\dagger
}-kQ\boldsymbol{\sigma}_{3}Q^{-1}\nonumber\\
&  ={\frac{1}{2r}}QPQ\boldsymbol{\sigma}_{3}P-{\frac{1}{2r}}P^{\dagger
}Q^{\dagger}Q\boldsymbol{\sigma}_{3}P-kQ\boldsymbol{\sigma}_{3}Q^{-1}%
\nonumber\\
&  =-P^{\dagger}\boldsymbol{\sigma}_{3}P+{\frac{1}{2r}}QPQ\boldsymbol{\sigma
}_{3}P-kQ\boldsymbol{\sigma}_{3}Q^{-1}. \label{RLga}%
\end{align}
Then, comparing (\ref{Ps3P+Qs3Q}) and (\ref{RLga}), we find that
\[
-\left(  P^{\dagger}\boldsymbol{\sigma}_{3}P+\lambda^{2}Q\boldsymbol{\sigma
}_{3}Q^{\dagger}\right)  =A,
\]
since
\[
\left(  P^{\dagger}\boldsymbol{\sigma}_{3}P+\lambda^{2}Q\boldsymbol{\sigma
}_{3}Q^{\dagger}\right)  ^{\dagger}=P^{\dagger}\boldsymbol{\sigma}%
_{3}P+\lambda^{2}Q\boldsymbol{\sigma}_{3}Q^{\dagger}.
\]
Namely, this fact implies that the RL vector appearing in the Kepler problem
in the 3 dimensional space is a remnant of the SO(4) moment map in the phase
space of the 4 dimensional space.

\section{Conclusion}

We have considered in this paper how RL vector in the Kepler system is related
to the moment map of the SO(4) group in the phase space of the 4 dimensional
particle motion, using the formalism of geometric algebra. We were able to
show that the RL vector is a representation of a moment map of an axial part
SO(3) connected to the SO(4) rotation that is KS transformed to 3 dimensional
space of the Kepler system. This explains the existence of RL vector which
apparently has nothing to do with the moment map. Furthermore, it is
interesting to note that, only for the closed orbit, i.e. for negative $h$, RL
vector appears as a remnant of the moment map.

Unlike other conserved quantities that are consequences of the geometric
symmetries of the system, conservation of RL vector has been considered to
have its origin in the dynamical properties of the Kepler
problem~\cite{Goldstein}. We have seen in this article that RL vector is also
a consequence of geometric symmetry of $\mathbb{R}^{4}\times\mathbb{R}^{4}$
phase space, which manifests itself only for the closed orbits of the
corresponding Kepler system. \vfill
\eject

\section*{Appendix A}

We show the $K_{i}$'s and $L_{i}$'s are generators of SO(4) rotation in the
$\mathbb{R}^{4}\times\mathbb{R}^{4}$ phase space. The Poisson brackets of
$K_{i}$'s and $L_{i}$'s for $(q_{a},\ p_{a})$ are
\begin{align*}
\{K_{1},q_{0}\}  &  ={\frac{1}{2}}\{J_{13}-J_{02},q_{0}\}={\frac{1}{2}}%
p_{2},\\
\{K_{1},q_{1}\}  &  ={\frac{1}{2}}\{J_{13}-J_{02},q_{1}\}=-{\frac{1}{2}}%
p_{3},\\
\{K_{1},q_{2}\}  &  ={\frac{1}{2}}\{J_{13}-J_{02},q_{2}\}={\frac{1}{2}}%
p_{0},\\
\{K_{1},q_{3}\}  &  ={\frac{1}{2}}\{J_{13}-J_{02},q_{3}\}=-{\frac{1}{2}}%
p_{1},\\
\{K_{1},p_{0}\}  &  ={\frac{1}{2}}\{J_{13}-J_{02},p_{0}\}=-{\frac{1}{2}}%
q_{2},\\
\{K_{1},p_{1}\}  &  ={\frac{1}{2}}\{J_{13}-J_{02},p_{1}\}={\frac{1}{2}}%
q_{3},\\
\{K_{1},p_{2}\}  &  ={\frac{1}{2}}\{J_{13}-J_{02},p_{2}\}=-{\frac{1}{2}}%
q_{0},\\
\{K_{1},p_{3}\}  &  ={\frac{1}{2}}\{J_{13}-J_{02},p_{3}\}={\frac{1}{2}}q_{1},
\end{align*}
\begin{align*}
\{K_{2},q_{0}\}  &  ={\frac{1}{2}}\{J_{01}+J_{23},q_{0}\}=-{\frac{1}{2}}%
p_{1},\\
\{K_{2},q_{1}\}  &  ={\frac{1}{2}}\{J_{01}+J_{23},q_{1}\}=-{\frac{1}{2}}%
p_{0},\\
\{K_{2},q_{2}\}  &  ={\frac{1}{2}}\{J_{01}+J_{23},q_{2}\}=-{\frac{1}{2}}%
p_{3},\\
\{K_{2},q_{3}\}  &  ={\frac{1}{2}}\{J_{01}+J_{23},q_{3}\}=-{\frac{1}{2}}%
p_{2},\\
\{K_{2},p_{0}\}  &  ={\frac{1}{2}}\{J_{01}+J_{23},p_{0}\}={\frac{1}{2}}%
q_{1},\\
\{K_{2},p_{1}\}  &  ={\frac{1}{2}}\{J_{01}+J_{23},p_{1}\}={\frac{1}{2}}%
q_{0},\\
\{K_{2},p_{2}\}  &  ={\frac{1}{2}}\{J_{01}+J_{23},p_{2}\}={\frac{1}{2}}%
q_{3},\\
\{K_{2},p_{3}\}  &  ={\frac{1}{2}}\{J_{01}+J_{23},p_{3}\}={\frac{1}{2}}q_{2},
\end{align*}
\begin{align*}
\{K_{3},q_{0}\}  &  ={\frac{1}{2}}\{J_{00}+J_{33}-J_{11}-J_{22},q_{0}%
\}=-{\frac{1}{2}}p_{0},\\
\{K_{3},q_{1}\}  &  ={\frac{1}{2}}\{J_{00}+J_{33}-J_{11}-J_{22},q_{1}%
\}={\frac{1}{2}}p_{1},\\
\{K_{3},q_{2}\}  &  ={\frac{1}{2}}\{J_{00}+J_{33}-J_{11}-J_{22},q_{2}%
\}={\frac{1}{2}}p_{2},\\
\{K_{3},q_{3}\}  &  ={\frac{1}{2}}\{J_{00}+J_{33}-J_{11}-J_{22},q_{3}%
\}=-{\frac{1}{2}}p_{3},\\
\{K_{3},p_{0}\}  &  ={\frac{1}{2}}\{J_{00}+J_{33}-J_{11}-J_{22},p_{0}%
\}={\frac{1}{2}}q_{0},\\
\{K_{3},p_{1}\}  &  ={\frac{1}{2}}\{J_{00}+J_{33}-J_{11}-J_{22},p_{1}%
\}=-{\frac{1}{2}}q_{1},\\
\{K_{3},p_{2}\}  &  ={\frac{1}{2}}\{J_{00}+J_{33}-J_{11}-J_{22},p_{2}%
\}=-{\frac{1}{2}}q_{2},\\
\{K_{3},p_{3}\}  &  ={\frac{1}{2}}\{J_{00}+J_{33}-J_{11}-J_{22},p_{3}%
\}={\frac{1}{2}}q_{3},
\end{align*}
\begin{align*}
\{L_{1},q_{0}\}  &  ={\frac{1}{2}}\{q_{1}p_{0}-q_{0}p_{1}+q_{2}p_{3}%
-q_{3}p_{2},q_{0}\}=-{\frac{1}{2}}q_{1},\\
\{L_{1},q_{1}\}  &  ={\frac{1}{2}}\{q_{1}p_{0}-q_{0}p_{1}+q_{2}p_{3}%
-q_{3}p_{2},q_{1}\}={\frac{1}{2}}q_{0},\\
\{L_{1},q_{2}\}  &  ={\frac{1}{2}}\{q_{1}p_{0}-q_{0}p_{1}+q_{2}p_{3}%
-q_{3}p_{2},q_{2}\}={\frac{1}{2}}q_{3},\\
\{L_{1},q_{3}\}  &  ={\frac{1}{2}}\{q_{1}p_{0}-q_{0}p_{1}+q_{2}p_{3}%
-q_{3}p_{2},q_{3}\}=-{\frac{1}{2}}q_{2},\\
\{L_{1},p_{0}\}  &  ={\frac{1}{2}}\{q_{1}p_{0}-q_{0}p_{1}+q_{2}p_{3}%
-q_{3}p_{2},p_{0}\}=-{\frac{1}{2}}p_{1},\\
\{L_{1},p_{1}\}  &  ={\frac{1}{2}}\{q_{1}p_{0}-q_{0}p_{1}+q_{2}p_{3}%
-q_{3}p_{2},p_{1}\}={\frac{1}{2}}p_{0},\\
\{L_{1},p_{2}\}  &  ={\frac{1}{2}}\{q_{1}p_{0}-q_{0}p_{1}+q_{2}p_{3}%
-q_{3}p_{2},p_{2}\}={\frac{1}{2}}p_{3},\\
\{L_{1},p_{3}\}  &  ={\frac{1}{2}}\{q_{1}p_{0}-q_{0}p_{1}+q_{2}p_{3}%
-q_{3}p_{2},p_{3}\}=-{\frac{1}{2}}p_{2},
\end{align*}
\begin{align*}
\{L_{2},q_{0}\}  &  ={\frac{1}{2}}\{q_{2}p_{0}-q_{0}p_{2}+q_{3}p_{1}%
-q_{1}p_{3},q_{0}\}=-{\frac{1}{2}}q_{2},\\
\{L_{2},q_{1}\}  &  ={\frac{1}{2}}\{q_{2}p_{0}-q_{0}p_{2}+q_{3}p_{1}%
-q_{1}p_{3},q_{1}\}=-{\frac{1}{2}}q_{3},\\
\{L_{2},q_{2}\}  &  ={\frac{1}{2}}\{q_{2}p_{0}-q_{0}p_{2}+q_{3}p_{1}%
-q_{1}p_{3},q_{2}\}={\frac{1}{2}}q_{0},\\
\{L_{2},q_{3}\}  &  ={\frac{1}{2}}\{q_{2}p_{0}-q_{0}p_{2}+q_{3}p_{1}%
-q_{1}p_{3},q_{3}\}={\frac{1}{2}}q_{1},\\
\{L_{2},p_{0}\}  &  ={\frac{1}{2}}\{q_{2}p_{0}-q_{0}p_{2}+q_{3}p_{1}%
-q_{1}p_{3},p_{0}\}=-{\frac{1}{2}}p_{2},\\
\{L_{2},p_{1}\}  &  ={\frac{1}{2}}\{q_{2}p_{0}-q_{0}p_{2}+q_{3}p_{1}%
-q_{1}p_{3},p_{1}\}=-{\frac{1}{2}}p_{3},\\
\{L_{2},p_{2}\}  &  ={\frac{1}{2}}\{q_{2}p_{0}-q_{0}p_{2}+q_{3}p_{1}%
-q_{1}p_{3},p_{2}\}={\frac{1}{2}}p_{0},\\
\{L_{2},p_{3}\}  &  ={\frac{1}{2}}\{q_{2}p_{0}-q_{0}p_{2}+q_{3}p_{1}%
-q_{1}p_{3},p_{3}\}={\frac{1}{2}}p_{1},
\end{align*}
\begin{align*}
\{L_{3},q_{0}\}  &  ={\frac{1}{2}}\{q_{3}p_{0}-q_{0}p_{2}+q_{1}p_{2}%
-q_{2}p_{1},q_{0}\}=-{\frac{1}{2}}q_{3},\\
\{L_{3},q_{1}\}  &  ={\frac{1}{2}}\{q_{3}p_{0}-q_{0}p_{2}+q_{1}p_{2}%
-q_{2}p_{1},q_{1}\}={\frac{1}{2}}q_{2},\\
\{L_{3},q_{2}\}  &  ={\frac{1}{2}}\{q_{3}p_{0}-q_{0}p_{2}+q_{1}p_{2}%
-q_{2}p_{1},q_{2}\}=-{\frac{1}{2}}q_{1},\\
\{L_{3},q_{3}\}  &  ={\frac{1}{2}}\{q_{3}p_{0}-q_{0}p_{2}+q_{1}p_{2}%
-q_{2}p_{1},q_{3}\}={\frac{1}{2}}q_{0},\\
\{L_{3},p_{0}\}  &  ={\frac{1}{2}}\{q_{3}p_{0}-q_{0}p_{2}+q_{1}p_{2}%
-q_{2}p_{1},p_{0}\}=-{\frac{1}{2}}p_{3},\\
\{L_{3},p_{1}\}  &  ={\frac{1}{2}}\{q_{3}p_{0}-q_{0}p_{2}+q_{1}p_{2}%
-q_{2}p_{1},p_{1}\}={\frac{1}{2}}p_{2},\\
\{L_{3},p_{2}\}  &  ={\frac{1}{2}}\{q_{3}p_{0}-q_{0}p_{2}+q_{1}p_{2}%
-q_{2}p_{1},p_{2}\}=-{\frac{1}{2}}p_{1},\\
\{L_{3},p_{3}\}  &  ={\frac{1}{2}}\{q_{3}p_{0}-q_{0}p_{2}+q_{1}p_{2}%
-q_{2}p_{1},p_{3}\}={\frac{1}{2}}p_{0}.
\end{align*}
These infinitesimal rotations are represented by the matrices $\Sigma_{i}$'s,
$\Lambda_{i}$'s.%

\begin{align*}
\Sigma_{1}\left(
\begin{array}
[c]{c}%
Q\\
P
\end{array}
\right)  ={\frac{1}{2}}\left(
\begin{array}
[c]{rrrr|rrrr}%
0 & 0 & 0 & 0 & 0 & 0 & 1 & 0\\
0 & 0 & 0 & 0 & 0 & 0 & 0 & -1\\
0 & 0 & 0 & 0 & 1 & 0 & 0 & 0\\
0 & 0 & 0 & 0 & 0 & -1 & 0 & 0\\\hline
0 & 0 & -1 & 0 & 0 & 0 & 0 & 0\\
0 & 0 & 0 & 1 & 0 & 0 & 0 & 0\\
-1 & 0 & 0 & 0 & 0 & 0 & 0 & 0\\
0 & 1 & 0 & 0 & 0 & 0 & 0 & 0
\end{array}
\right)  \left(
\begin{array}
[c]{c}%
q_{0}\\
q_{1}\\
q_{2}\\
q_{3}\\
p_{0}\\
p_{1}\\
p_{2}\\
p_{3}%
\end{array}
\right)  ={\frac{1}{2}} \left(
\begin{array}
[c]{r}%
p_{2}\\
-p_{3}\\
p_{0}\\
-p_{1}\\
-q_{2}\\
q_{3}\\
-q_{0}\\
q_{1}%
\end{array}
\right)  ,
\end{align*}
\begin{align*}
\Sigma_{2}\left(
\begin{array}
[c]{c}%
Q\\
P
\end{array}
\right)  ={\frac{1}{2}}\left(
\begin{array}
[c]{rrrr|rrrr}%
0 & 0 & 0 & 0 & 0 & -1 & 0 & 0\\
0 & 0 & 0 & 0 & -1 & 0 & 0 & 0\\
0 & 0 & 0 & 0 & 0 & 0 & 0 & -1\\
0 & 0 & 0 & 0 & 0 & 0 & -1 & 0\\\hline
0 & 1 & 0 & 0 & 0 & 0 & 0 & 0\\
1 & 0 & 0 & 0 & 0 & 0 & 0 & 0\\
0 & 0 & 0 & 1 & 0 & 0 & 0 & 0\\
0 & 0 & 1 & 0 & 0 & 0 & 0 & 0
\end{array}
\right)  \left(
\begin{array}
[c]{c}%
q_{0}\\
q_{1}\\
q_{2}\\
q_{3}\\
p_{0}\\
p_{1}\\
p_{2}\\
p_{3}%
\end{array}
\right)  ={\frac{1}{2}} \left(
\begin{array}
[c]{r}%
p_{1}\\
p_{0}\\
p_{3}\\
p_{2}\\
q_{1}\\
q_{0}\\
q_{3}\\
q_{2}%
\end{array}
\right)  ,
\end{align*}
\begin{align*}
\Sigma_{3}\left(
\begin{array}
[c]{c}%
Q\\
P
\end{array}
\right)  ={\frac{1}{2}}\left(
\begin{array}
[c]{rrrr|rrrr}%
0 & 0 & 0 & 0 & -1 & 0 & 0 & 0\\
0 & 0 & 0 & 0 & 0 & 1 & 0 & 0\\
0 & 0 & 0 & 0 & 0 & 0 & 1 & 0\\
0 & 0 & 0 & 0 & 0 & 0 & 0 & -1\\\hline
1 & 0 & 0 & 0 & 0 & 0 & 0 & 0\\
0 & -1 & 0 & 0 & 0 & 0 & 0 & 0\\
0 & 0 & -1 & 0 & 0 & 0 & 0 & 0\\
0 & 0 & 0 & 1 & 0 & 0 & 0 & 0
\end{array}
\right)  \left(
\begin{array}
[c]{c}%
q_{0}\\
q_{1}\\
q_{2}\\
q_{3}\\
p_{0}\\
p_{1}\\
p_{2}\\
p_{3}%
\end{array}
\right)  ={\frac{1}{2}} \left(
\begin{array}
[c]{r}%
-p_{0}\\
p_{1}\\
p_{2}\\
-p_{3}\\
q_{0}\\
-q_{1}\\
-q_{2}\\
q_{3}%
\end{array}
\right)  ,
\end{align*}
\begin{align*}
\Lambda_{1}\left(
\begin{array}
[c]{c}%
Q\\
P
\end{array}
\right)  ={\frac{1}{2}}\left(
\begin{array}
[c]{rrrr|rrrr}%
0 & -1 & 0 & 0 & 0 & 0 & 0 & 0\\
1 & 0 & 0 & 0 & 0 & 0 & 0 & 0\\
0 & 0 & 0 & 1 & 0 & 0 & 0 & 0\\
0 & 0 & -1 & 0 & 0 & 0 & 0 & 0\\\hline
0 & 0 & 0 & 0 & 0 & -1 & 0 & 0\\
0 & 0 & 0 & 0 & 1 & 0 & 0 & 0\\
0 & 0 & 0 & 0 & 0 & 0 & 0 & 1\\
0 & 0 & 0 & 0 & 0 & 0 & -1 & 0
\end{array}
\right)  \left(
\begin{array}
[c]{c}%
q_{0}\\
q_{1}\\
q_{2}\\
q_{3}\\
p_{0}\\
p_{1}\\
p_{2}\\
p_{3}%
\end{array}
\right)  ={\frac{1}{2}} \left(
\begin{array}
[c]{r}%
-q_{1}\\
q_{0}\\
q_{3}\\
-q_{2}\\
-p_{1}\\
p_{0}\\
p_{3}\\
-p_{2}%
\end{array}
\right)  ,
\end{align*}
\begin{align*}
\Lambda_{2}\left(
\begin{array}
[c]{c}%
Q\\
P
\end{array}
\right)  ={\frac{1}{2}}\left(
\begin{array}
[c]{rrrr|rrrr}%
0 & 0 & -1 & 0 & 0 & 0 & 0 & 0\\
0 & 0 & 0 & -1 & 0 & 0 & 0 & 0\\
1 & 0 & 0 & 0 & 0 & 0 & 0 & 0\\
0 & 1 & 0 & 0 & 0 & 0 & 0 & 0\\\hline
0 & 0 & 0 & 0 & 0 & 0 & -1 & 0\\
0 & 0 & 0 & 0 & 0 & 0 & 0 & -1\\
0 & 0 & 0 & 0 & 1 & 0 & 0 & 0\\
0 & 0 & 0 & 0 & 0 & 1 & 0 & 0
\end{array}
\right)  \left(
\begin{array}
[c]{c}%
q_{0}\\
q_{1}\\
q_{2}\\
q_{3}\\
p_{0}\\
p_{1}\\
p_{2}\\
p_{3}%
\end{array}
\right)  ={\frac{1}{2}} \left(
\begin{array}
[c]{r}%
-q_{2}\\
-q_{3}\\
q_{0}\\
q_{1}\\
-p_{2}\\
-p_{3}\\
p_{0}\\
p_{1}%
\end{array}
\right)  ,
\end{align*}
\begin{align*}
\Lambda_{3}\left(
\begin{array}
[c]{c}%
Q\\
P
\end{array}
\right)  ={\frac{1}{2}}\left(
\begin{array}
[c]{rrrr|rrrr}%
0 & 0 & 0 & -1 & 0 & 0 & 0 & 0\\
0 & 0 & 1 & 0 & 0 & 0 & 0 & 0\\
0 & -1 & 0 & 0 & 0 & 0 & 0 & 0\\
1 & 0 & 0 & 0 & 0 & 0 & 0 & 0\\\hline
0 & 0 & 0 & 0 & 0 & 0 & 0 & -1\\
0 & 0 & 0 & 0 & 0 & 0 & 1 & 0\\
0 & 0 & 0 & 0 & 0 & -1 & 0 & 0\\
0 & 0 & 0 & 0 & 1 & 0 & 0 & 0
\end{array}
\right)  \left(
\begin{array}
[c]{c}%
q_{0}\\
q_{1}\\
q_{2}\\
q_{3}\\
p_{0}\\
p_{1}\\
p_{2}\\
p_{3}%
\end{array}
\right)  ={\frac{1}{2}} \left(
\begin{array}
[c]{r}%
-q_{3}\\
q_{2}\\
-q_{1}\\
q_{0}\\
-p_{3}\\
p_{2}\\
-p_{1}\\
p_{0}%
\end{array}
\right)  .
\end{align*}
\vfill
\eject

\section*{Appendix B}

In the Euclidean 3 dimensional space, a vector $\boldsymbol{a}$ is expressed
as
\[
\boldsymbol{a}=a_{1}\boldsymbol{\sigma}_{1}+a_{2}\boldsymbol{\sigma}_{2}%
+a_{3}\boldsymbol{\sigma}_{3},
\]
which is called an object of grade 1 in the geometric algebra, since
$\boldsymbol{a}$ has terms proportional to $\boldsymbol{\sigma}_{i}$. By
virture of the relation (\ref{GA}), geometric product of arbitrary two vectors
$\boldsymbol{a}$ and $\boldsymbol{b}$ can be written as
\begin{align*}
\boldsymbol{ab}=  &  a_{1}b_{1}+a_{2}b_{2}+a_{3}b_{3}\\
&  +(a_{2}b_{3}-a_{3}b_{2})\boldsymbol{\sigma}_{2}\boldsymbol{\sigma}%
_{3}+(a_{3}b_{1}-a_{1}b_{3})\boldsymbol{\sigma}_{3}\boldsymbol{\sigma}%
_{1}+(a_{1}b_{2}-a_{2}b_{1})\boldsymbol{\sigma}_{1}\boldsymbol{\sigma}_{2}\\
=  &  \boldsymbol{a}\cdot\boldsymbol{b}+\boldsymbol{a}\times\boldsymbol{b},
\end{align*}
where scalar part $\boldsymbol{a}\cdot\boldsymbol{b}$ has grade 0, and
bivector part $\boldsymbol{a}\times\boldsymbol{b}$ is grade 2. A multivector
$A$ may be decomposed into the grade-projection operator $\langle A\rangle
_{i}$. The most general multivector is expressed as
\[
A=\langle A\rangle_{0}+\langle A\rangle_{1}+\langle A\rangle_{2}+\langle
A\rangle_{3}%
\]
In the 3 dimensional Euclidean space, the highest grade of the algebra is
grade 3. The term of the grade 3, i.e., $\langle A\rangle_{3}$ is proportional
to $I$. Using $\boldsymbol{\sigma}_{i}$, $I$ is expressed as
\[
I=\boldsymbol{\sigma}_{1}\boldsymbol{\sigma}_{2}\boldsymbol{\sigma}_{3}.
\]
The grade with more than of 3 does not exist, since multiplying $I$ by
$\boldsymbol{\sigma}_{i}$ gives grade 2. $I$ has a property similar to the
imaginary unit, $I^{2}=-1$.

\section*{Acknowledgment}

We would like to thank Akihiro Nakayama for discussions, support and
hospitality. \vfill
\eject

\end{document}